\documentclass[12pt,a4paper]{article}
\usepackage{mathtools}
\usepackage{amsfonts,amssymb,slashed,latexsym,amsmath,multirow,color,,epsf,dsfont,graphics,graphicx}

\newcommand{\be}{\begin{equation}}
\newcommand{\ee}{\end{equation}}
\newcommand{\nnr}{\nonumber \\}

\newcommand{\fr}{\frac}

\newcommand{\df}{d}
\newcommand{\expe}[1]{\textrm{e}^{#1}}

\newcommand{\sr}{\sqrt}

\newcommand{\im}{\textrm{i}}
\def\bea{\begin{eqnarray}}
\def\eea{\end{eqnarray}}
\newcommand{\ft}[2]{{\textstyle\frac{#1}{#2}}}

\def\a{\alpha}
\def\b{\beta}
\def\g{\gamma}

\def\d{\delta}

\def\e{\epsilon}

\def\k{\kappa}
\def\l{\lambda}

\def\m{\mu}
\def\n{\nu}
\def\r{\rho}
\def\s{\sigma}

\def\t{\tau}

\def\o{\omega}

\makeatletter \@addtoreset{equation}{section} \makeatother

\begin{document}
	\begin{titlepage}
		\begin{center}
			\vskip 1.5cm
			
			{\Large \bf Rotating Strings in Six-Dimensional Higher-Derivative Supergravity}
			
			\vskip 2cm
			
			{\bf David D. K. Chow and Yi Pang${}^1$} \\
			
			\vskip 15pt

			{\em $^1$ \hskip -.1truecm
				\em Mathematical Institute, University of Oxford, \\
				Woodstock Road, Oxford OX2 6GG, U.K. \vskip 5pt }
			
			{email: {\tt Yi.Pang@maths.ox.ac.uk}} \\
			
			\vskip .4truecm

		\end{center}
		
		\vskip 0.5cm
		
		\begin{center}
			{\bf ABSTRACT}\\[3ex]
		\end{center}
		
			We construct the first analytical rotating string solution in 6-dimensional Einstein--Gauss--Bonnet supergravity, carrying both electric and magnetic charges. By embedding the known rotating string solution of the 2-derivative theory into 6-dimensional off-shell supergravity, the Killing spinors associated with the underlying supersymmetry can be made off-shell and are universal to all off-shell supergravity models based on the same field content. The near-horizon geometry is S$^3$ fibred over the extremal BTZ black hole, locally isomorphic to AdS$_3\times$S$^3$.  We compute the higher-derivative corrections to the Brown--Henneaux central charges in a particular $R+R^2$ model resulting from K3 compactification of type IIA string theory.
		
	\end{titlepage}
	
	\newpage
	
	
	\section{Introduction}

Although Einstein's theory of general relativity (GR) is well-established, it is unlikely to be completed at ultra-violet energy scales on its own. This is because, in a similar manner to Fermi's effective theory for weak interactions, Newton's constant characterizing the strength of the gravitational coupling carries units of length-squared. The formation of black holes at Planck scale energy density and the concurrence  of space-time singularities further obscure the nature of gravity at short distances, where quantum effects becomes important.

To unify gravity and quantum mechanics, string theory predicts that GR should be augmented with new degrees of freedom in such a way that supersymmetry is respected at sufficiently high energy scales. In perturbative string theory, quantum corrections to GR can in principle be calculated at each string loop order, in which taking the low energy limit yields an infinite series expanded in powers of the string length squared $\ell_s^2$, with supersymmetry preserved order by order in $\ell_s^2$. Four-dimensional supergravity models preserving eight supercharges arise as low energy limits of string theory compactified on Calabi--Yau 3-folds. The leading higher-derivative corrections to the Einstein--Hilbert action contain terms quadratic in the curvature tensor coupled to scalar fields.  

The recent observations of gravitational waves (GW) from binary black holes \cite{GW150914} and neutron stars \cite{GW170817} mergers have extended the success of GR from the low-velocity, weak gravitational field regime to a highly dynamical regime governed by strong gravity. Interestingly, effects from string-inspired Einstein--dilaton--Gauss--Bonnet and dynamical Chern--Simons gravity on gravitational wave emission have been seriously considered in exploring modifications to GR, as they result in the emission of scalar dipole (at 1PN) and quadrupole (at 2PN) radiation during the inspiral \cite{Nair:2019iur}. This could potentially revive interest in higher-derivative extended gravity models with a string theory origin with the advantage that, given a compactification scheme, the lower dimension couplings are fixed, thus increasing verifiability of the model.

   This paper is devoted to understanding exact solutions in higher-derivative extended supergravity models, which are notoriously hard.  Supersymmetric higher-derivative terms can be most conveniently organized using the off-shell formulation of supersymmetry, which is possible up to six dimensions ($D=6$), in which we set up the calculation. By consistent dimensional reduction, six-dimensional solutions give rise to solutions in lower-dimensional models. This leaves out a separate case in $D=5$ (and of course its reduction to lower dimensions) due to the possibility of introducing two inequivalent off-shell formulations \cite{Bergshoeff:2001hc}, which we leave for a future study. The main advantage of working in $D=6$ is that there are fewer fields, and so correspondingly the supersymmetry transformation rules are reasonably simple. For on-shell two-derivative supergravities in $D = 6$, there are classifications of supersymmetric solutions in \cite{GMR, CC, Cano:2018wnq, LV}. In this paper the approach uses the crucial fact that the off-shell supersymmetry transformation rules are theory-independent. Therefore, solutions to off-shell Killing spinor equations are universal to all theories invariant under the same supersymmetry transformations. One can thus use the known explicit supersymmetric solutions of the simplest two-derivative theory as a guide to deduce the structure of the off-shell Killing spinor applicable to other solutions with the same isometries. Substituting the ansatz for bosonic fields based on symmetry considerations and
   the universal form of the off-shell Killing spinor into the off-shell Killing spinor equations results in
   theory-independent algebraic relations amongst the undetermined functions. Therefore,
   without using equations of motion, we have partially solved the system for all supergravity models. Of course, fully determining the solutions requires dynamical input from a specific theory. For simplicity, we
   choose the $p$-form field equations as the dynamical input, which usually admit a first integral implied by the gauge symmetry. We elaborate this idea by constructing the supersymmetric, rotating, dyonic string solution in $D=6$ ungauged minimal supergravity coupled to a tensor multiplet with higher-derivative extensions. The static and rotating dyonic strings solutions in $D=6$ two-derivative supergravity were presented in  \cite{Duff:1993ye,DFKR,DKL,CL}.

   The near-horizon limit of the dyonic string solution leads to the conformally flat AdS$_3\times$S$^3$ vacuum preserving eight chiral supercharges. The central charges present in the asymptotic symmetry algebra of AdS$_3$ receive higher-derivative corrections. When the supergravity model arises from string theory, the AdS$_3$ vacuum admits a dual 2D conformal field theory (CFT) description. Comparing the higher-derivative corrected gravitational central charges to the CFT central charges then provides a precision test of the AdS$_3$/CFT$_2$ correspondence beyond the large central charge limit. Using our solution, we compute the higher derivative corrections to the Brown--Henneaux central charges in a particular model resulting from compactifying type IIA string theory on a K3 manifold.

\section{Rotating dyonic string solutions}

\subsection{Supergravity}

We begin by introducing the $D=6$ off-shell formulation of supergravity. The minimal supergravity multiplet combines with a self-dual tensor multiplet to form the off-shell irreducible dilaton--Weyl multiplet consisting of bosonic fields $\{g_{\m\n}\,,B_{\m\n}\,,L\,,V_{\m}\,,V^{ij}_\m\,,E_\m\}$ and fermionic fields $\{\psi^i_\m\,,\varphi^i\} $, with $i=1,2$ and $\m$ labeling the 6D coordinate index. $\{V_{\m}\,,V^{ij}_\m\,,E_\m\}$ are the auxiliary fields required by the off-shell closure of the supersymmetry algebra. The Einstein--Hilbert Lagrangian invariant under the 6D off-shell supersymmetry is given by
 \be
\label{EHSUGRA}
{\cal L}_{\rm EH}=\sr{-g}\left(LR +L^{-1}\nabla^\mu L \, \nabla_\mu L - \ft{1}{12}L H^{\mu \nu \rho} H_{\mu \nu \rho}+\ldots\right)\,,\quad H_{\m\n\r}=3\partial_{[\m} B_{\n\r]}\,,
\ee
where here and below the omitted terms are quadratic in auxiliary fields. This Lagrangian is written in string frame, which simplifies the supersymmetrization of higher-derivative terms. In addition to \eqref{EHSUGRA}, we add the recently constructed off-shell supersymmetric Gauss--Bonnet term \cite{NOPT}\footnote{Here we use the convention $\epsilon_{012345}=1$, so the $B\wedge R\wedge R$ term has a minus sign, whereas \cite{NOPT} has a plus sign. This sign difference results from a different convention for the chirality of fermions.}
\bea
\label{GBSUGRA}
{\cal L}_{\rm GB} &=& \sr{-g} \left(  R_{\m\n\r\s} R^{\m\n\r\s} - 4 R_{\m\n} R^{\m\n} + R^2 + \ft16 R H^2 - R^{\m\n} H_{\m\n}^2 + \ft12 R_{\m\n\r\s} H^{\m\n\l} H^{\r\s}{}_{\l} \right. \nnr
&& \left. + \ft{5}{24} H^4 + \ft{1}{144} (H^2)^2 - \ft18 (H_{\m\n}^2)^2 - \ft14 \e^{\m\n\r\s\l\t} B_{\m\n} R_{\r\s}{}^{\a}{}_{\b} (\o_+) R_{\l\t}{}^{\b}{}_{\a}(\o_+)+\ldots \right)\,,
\eea
where $R_{\m\n\r\s}$ is the standard Riemann tensor of $g_{\m\n}$, while $R_{\m\n}{}^{\a}{}_{\b} (\o_+)$ is the curvature associated with the torsionful spin connection $\o_{+\m}^{\a}{}_{\b}$,
\be
R_{\m\n}{}^{\a}{}_{\b} (\o_+)=\partial_{\m}\o_{+\n}^{\a}{}_{\b}+\o_{+\m}^{\a}{}_{\g}\,\o_{+\n}^{\g}{}_{\b}-(\m\leftrightarrow\n)\,,\quad \o_{+\m}^\a{}_\b=\o_{\m}^\a{}_\b+\ft12H_{\m}{}^\a{}_\b\,.
\ee
The shorthand notations for various contractions of $H_{\m\n\r}$ are
\be
H^2:=H_{\m\n\r}H^{\m\n\r}\,,\quad H_{\m\n}^2=H_{\m\r\s}H_{\n}{}^{\r\s}\,,\quad H^4:=H_{\m\n\s}H_{\r\l}{}^{\s}H^{\m\r\d}H^{\n\l}{}_{\d}\,.
\ee
Other 6D off-shell supersymmetric higher-derivative actions can be found in \cite{BSS86}-\cite{BNOPT}. We prefer the Gauss--Bonnet combination to other higher-derivative invariants because it gives rise to relatively simple equations of motion while entailing enough complexity to display the efficiency of our method, especially when applied to non-static solutions.  Furthermore, going to Einstein frame, the model becomes the supersymmetric version of the Einstein--dilaton--Gauss--Bonnet gravity of broader interest. To summarize, in this section, we use the total action
\be
S_{\rm EGB}=\frac1{16\pi G_6}\int d^6 x \left({\cal L}_{\rm EH}+\frac{1}{16}\l_{\rm GB}\,{\cal L}_{\rm GB}\right) ,
\ee
where $G_6$ is the $D = 6$ Newton constant.  The action enjoys the scaling symmetry
\be
(g_{\m\n}\,,B_{\m\n}\,,L\,,\l_{\rm GB}) \rightarrow (tg_{\m\n}\,,tB_{\m\n}\,,t^{-2}L\,,t^{-1}\l_{\rm GB} )
\label{scaling}
\ee
needed in later discussions.

\subsection{Killing spinor equation}

The supersymmetric rotating string solutions can be derived from the 3-charge black string solution \cite{CL} by taking the limit in which the mass saturates the Bogomolny--Prasad--Sommerfield (BPS) bound set by the sum of all charges. Also, the two angular momenta are equal and the solution possesses an enhanced U(2) isometry \cite{GLPP}.
The explicit form of the solution is given by
\bea
\label{CLs}
ds_6^2&=&\frac{r^2}{(r^2+Q)}\bigg[-\bigg( dt+\frac{J}{2r^2}\s_3 \bigg) ^2+ \bigg( dx+\frac{J}{2r^2}\s_3 \bigg) ^2 \bigg]
+(r^2+P)\frac{dr^2}{r^2}\nnr
&&+\ft14(r^2+P)\left(\s_3^2+d\theta^2+\sin^2\theta d\phi^2\right)\,,\nnr
B_{(2)}&=&2P\o_2+\frac{r^2}{r^2+Q}dt\wedge dx+\frac{J}{2(r^2+Q)}dt\wedge\s_3+\frac{J}{2(r^2+Q)}\s_3\wedge dx\,,\nnr
L&=&\frac{r^2+Q}{r^2+P}\,,
\eea
where $\s_3=d\chi-\cos\theta d\phi$, $d\o_2=\textrm{Vol}(S^3)$, and the electromagnetic charges $Q$ and $P$ are positive. All auxiliary fields vanish for this solution. It is evident that the solution above is cohomogeneity-1 and admits a U(2)$\times\mathbb{R}^2$ isometry. The infinite string lies along the $x$-direction. When $r\to\infty$, the solution approaches Minkowski spacetime with energy and angular-momentum density along the string
\be
{\cal M}=\frac{(Q+P)\pi}{4 G_6}\,,\qquad {\cal J}=-\frac{J\pi}{4 G_6}\,,
\ee
where the angular momentum is associated with the Euler angle $\chi$. In the asymptotically flat case, the symmetry is given by the Poincar\'{e} superalgebra not involving the generators associated compact isometry of the solution. Therefore, the angular momentum does not enter the BPS bound. The supersymmetry of the solution \eqref{CLs} can be seen by explicitly solving the Killing spinor equations, which are given by the vanishing supersymmetry variations of the fermionic fields $\{\psi^i_\m\,,\varphi^i\}$:
\begin{align}
\label{susyt}
& \bigg( \partial_{\mu} +\frac14\omega_{\mu
	\a\b}\gamma^{\a\b} \bigg) \epsilon^i-\frac12{V}_\m\d^{ij}\epsilon_j +{V}_\mu{}^i{}_j\epsilon^j +\frac18 H_{\mu\nu\rho}\gamma^{\nu\rho}\epsilon^i = 0 \, ,
\nnr
& \frac{1}{2\sqrt 2} \gamma^\mu
\delta^{ij}\partial_\mu L \epsilon_j -\frac14 \gamma^\mu E_\mu\epsilon^i
+\frac{1}{\sqrt2}\gamma^\mu { V}{}_\mu^{(i}{}_k \delta^{j)k} L
\epsilon_j - \frac{1}{12\sqrt2}L\delta^{ij}\gamma_{\m\n\r} H^{\m\n\r} \epsilon_j = 0 \,,
\end{align}
where $\epsilon^i$ is a symplectic Majorana--Weyl spinor with the chirality  $\g_{012345}\epsilon^i=\epsilon^i$.  Indices $i, j, \ldots$ are raised and lowered by $\varepsilon^{ij}$ and $\varepsilon_{ij}$. For convenience, we introduce the
complex Weyl spinor
\be
\epsilon=\epsilon_1+\im\epsilon_2.
\ee
Substituting \eqref{CLs} into \eqref{susyt}, we find the solution for $\epsilon$ given by
\be
\label{ks1}
\epsilon=\frac{r^{1/2}}{(r^2+Q)^{1/4}}\epsilon_0\,,
\ee
where $\epsilon_0$ is the Killing spinor on a two-sphere embedded in the $D=6$ spinor so that it satisfies the projection conditions
\be
\g^{012345}\epsilon_0=-\epsilon_0\,,\quad \g^{01}\epsilon_0=-\epsilon_0\,.
\ee
We now make an ansatz for supersymmetric rotating strings in generic supergravity models based on the same set of fields. The metric $g_{\m\n}$, two-form potential $B_{\m\n}$ and the scalar $L$ takes the form fixed by the U(2)$\times\mathbb{R}^2$ isometry
\begin{align}
\label{genstr}
ds_6^2&=a^2(r)\left[-(dt+\varpi\s_3)^2+(dx+\varpi\s_3)^2\right]
+b^2(r){dr^2}+\ft14c^2(r)\left(\s_3^2+d\theta^2+\sin^2\theta d\phi^2\right)\,,\nnr
B_{(2)}&=2P\o_2+d(r)dt\wedge dx+f(r)(dt - dx) \wedge\s_3 \,,\nnr
L&=L(r)\,,
\end{align}
with vanishing auxiliary fields. The corresponding Killing spinor is assumed to be
\be
\label{ks2}
\epsilon=\Pi(r)\epsilon_0\,,
\ee
where $\epsilon_0$ is the same as in \eqref{ks1}. Substituting \eqref{genstr} and \eqref{ks2} into the Killing spinor equations \eqref{susyt} then yields the consistency relations
\bea
\label{ks3}
&&\dot{d}-2 a \dot{a}=0\,,\quad \frac{\dot{d}}{a^2b}+\frac{\dot{L}}{Lb}-\frac{2 P}{c^3}=0\,, \quad \varpi \dot{d}-\dot{f}+a^2 \dot{\varpi}+\frac{2 f b}{c}-\frac{2 a^2b \varpi}{c}=0\,,\nnr
&& \frac{P b}{c} -bc+c\dot{c}=0\,,\quad \Pi=\sr{a}\,,
\eea
where dots denote derivatives with respect to $r$. The conditions in \eqref{ks3} are necessary and sufficient conditions for the existence of Killing spinors corresponding to the rotating strings. The gauge symmetries in the system are related to the choice of coordinate $r$ and the Abelian one-form symmetry of $B_{(2)}$. We can fix the reparametrization of $r$ by demanding
\be
c=rb\,,
\ee
using which \eqref{ks3} can be readily integrated yielding
\be
d=a^2+\k_1\,,\quad c^2=r^2+P\,,\quad \varpi d=f+\k_2 r^2\,,\quad L|d|=\frac{\k_3r^2}{r^2+P}\,.
\ee
We then set $\k_1=0$ using the Abelian one-form symmetry and $\k_3=1$ using the scaling \eqref{scaling}. We consider the case in which the higher-derivative correction is small and the general solution obeys the same boundary condition as indicated by solution
\eqref{CLs}. Explicitly,
\be
\label{bc}
a = {\cal O}(1)\,,\quad b = {\cal O}(1)\,,\quad c = {\cal O}(r)\,,\quad \varpi = {\cal O}(r^{-2})\,,\quad d = {\cal O}(1)\,,\quad f= {\cal O}(r^{-2})\,,
\ee
which implies that $\k_2=0$. In summary, we have
\be
\label{sol1}
a^2=d\,,\quad b^2=1+\frac{P}{r^2}\,,\quad c^2=r^2+P\,,\quad f=d\varpi\,,\quad L=\frac{r^2}{d(r^2+P)}\,,
\ee
so have reduced the system to solving for two functions, $d(r)$ and $\varpi(r)$.  We emphasize that these relations are independent of the details of the Lagrangian, depending only on the set of fields in the model and the U(2)$\times \mathbb{R}^2$ symmetry of the solution.

The dynamical equations determining $d(r)$ and $\varpi(r)$, and therefore the whole system, come from the components of the $B_{(2)}$ field equation. The dilaton equation is automatically satisfied for the off-shell relations \eqref{sol1}. There are two independent equations from the $B_{(2)}$ field equation. In terms of new variables
\be
\r=r^2\,,\quad \widetilde{\varpi}=r^2 \varpi\,,
\ee
one of them can be written as
\be
\left(\frac{(\widetilde{\varpi}-\l_{\rm GB} \widetilde{\varpi} d')'}{\r+P}\right)'=0\,,
\ee
where primes denote derivatives with respect to $\r$.  The boundary condition of \eqref{bc} leads to
\be
\widetilde{\varpi}=\frac{J}{2(1-\l_{\rm GB} d')}\,.
\ee
We have chosen the integration constant to agree with \eqref{CLs} in the $\l_{\rm GB}\to 0$ limit. Using conservation of electric charge, the equation for $d$ can also be easily derived. We find it can be integrated once more and leads to
\be
(Q+\r)d-\r+\frac{\l_{\rm GB}}2 \left[\frac{2P\r d'}{P+\r}-\frac{P^2d}{(P+\r)^2}\right]=0\, .
\label{deqn}
\ee
We see that the system has been reduced to a single ordinary differential equation for $d(\r)$.  Note that, up to a trivial overall factor, there is a scaling symmetry under
\be
(\lambda_{\rm GB} , Q, P, \rho, d) \rightarrow (t \lambda_{\rm GB} , t Q, t P, t \rho, d) ,
\ee
for constant $t$.  In the purely electric case $P=0$ the contribution from the Gauss--Bonnet invariant vanishes. However, in this case the solution is singular at $\r=0$, so we assume that $P > 0$ in the discussion below.  To explicitly solve \eqref{deqn}, we can rewrite it as
\be
\lambda_{\rm GB} P (\rho^{Q/\lambda_{\rm GB} - 1/2} \sr{\rho + P} \expe{(\rho + Q + P)^2/2 P \lambda_{\rm GB}} d(\rho))' = \rho^{Q/\lambda_{\rm GB} - 1/2} (\rho + P)^{3/2} \expe{(\rho + Q + P)^2/2 P \lambda_{\rm GB}} ,
\ee
and so the general solution for $d(\r)$ can be expressed as a integral. If there is a horizon at $\rho = \rho_0>0$, then $d(\rho_0) = 0$ and the solution exhibits a curvature singularity at $\r=\r_0$. At the same location, the dilaton field $L$ also diverges. Therefore, for regular solutions, the lower end of $\r$ can always reach 0. We will not consider negative $\r$ for the positivity of the dilaton field (string coupling constant). As \eqref{deqn} is a linear ODE with a source, its solution can be expressed a linear combination of the solution to the homogeneous equation and the special solution dictated by the source. Near $\r=0$, the solution is approximated by
\be
d(\r)\sim \r^{\frac{1}{2}-\frac{Q}{\lambda_{\rm GB} }}\left( c+\frac{2\r^{\frac{1}{2}+\frac{Q}{\lambda_{\rm GB} }}}{2Q+\lambda_{\rm GB}} \right)\,,
\label{appsol}
\ee
from which we can see when $Q/\lambda_{\rm GB} > -1/2$, the first term in \eqref{appsol} dominates and signals a curvature singularity at $\r=0$, unless we set the integration constant $c$ to 0. On the other hand, when $Q/\lambda_{\rm GB} < -1/2$, the second term in \eqref{appsol} dominates and the solution approaches an AdS$_3$ throat. In summary, imposing regularity of the solution, we have 
\be
d(\rho) \sim \fr{\rho}{Q + \lambda_{\rm GB}/2}\,.
\ee
This leads to the horizon value of the dilaton field given by
\be
L\vert_H=\frac{Q + \lambda_{\rm GB}/2}{P}\,,
\ee
whose positivity constrains 
\be
Q + \lambda_{\rm GB}/2>0\,.
\label{c1}
\ee
The overlap between \eqref{c1} and $Q/\lambda_{\rm GB} > -1/2$ implies $\l_{\rm GB}\ge0$. In this case, the exact solution can be expressed as
\be
d(\rho) = \frac{\rho ^{1/2 - Q/\lambda_{\rm GB}} \expe{- (\rho + Q + P)^2/2 \lambda_{\rm GB}  P}}{\lambda_{\rm GB} P \sqrt{P+\rho }}\int_0^\r \! \df y \, y^{Q/\lambda_{\rm GB} - 1/2} (y + P)^{3/2} \expe{(y + Q + P)^2 /2 \lambda_{\rm GB} P} .
\label{sol}
\ee
In the other case, \eqref{c1} and $Q/\lambda_{\rm GB} < -1/2$ can be simultaneously satisfied for negative $\lambda_{\rm GB}$ whose magnitude is bounded above by $2Q$. In this case, the exact solution can be expressed as
\be
d(\rho) =-\frac{\rho ^{1/2 - Q/\lambda_{\rm GB}} \expe{- (\rho + Q + P)^2/2 \lambda_{\rm GB}  P}}{\lambda_{\rm GB} P \sqrt{P+\rho }}\int_\r^{\infty} \! \df y \, y^{Q/\lambda_{\rm GB} - 1/2} (y + P)^{3/2} \expe{(y + Q + P)^2 /2 \lambda_{\rm GB} P} .
\label{sol}
\ee 
In both cases, $d(\rho) \rightarrow 1$ as $\rho \rightarrow \infty$, and in particular
\be
d(\r)=1-\frac{Q}{\r}+\frac{Q^2}{\r^2}+\cdots\,,
\ee
which implies that the conserved charges measured at infinity are not modified by the inclusion of the Gauss--Bonnet combination. 

\section{AdS$_3$/CFT$_2$}

Upon taking the $r\to 0$ near-horizon limit, the rotating dyonic string solution becomes a locally AdS$_3\times$S$^3$ background supported by an anti-self-dual three-form flux and a constant dilaton \cite{CL}. Local coordinate transformation brings the AdS$_3$ to the standard Ba\~{n}ados--Teitelboim--Zanelli (BTZ) black hole metric. We thus have BTZ$\times$S$^3$ as an exact solution to the $D=6$ supergravity extended by the Gauss--Bonnet combination,
\begin{align}
\label{NHS}
ds^2_6&=-\frac{(r^2-r^2_+)(r^2-r_-^2)}{\ell^2r^2}d\tau^2+r^2 \bigg( dy-\frac{r_-r_+}{\ell r^2}d\tau \bigg) ^2+\frac{\ell^2 r^2}{(r^2-r^2_+)(r^2-r_-^2)}dr^2+\ell^2ds_{{\rm S}^3}^2\,,\nnr
H_{(3)}&=2\ell^2\left(\Omega_{{\rm AdS}_3}+\Omega_{{\rm S}^3}\right)\,,\qquad L=L_0\,,\qquad \ell^2=P\,,
\end{align}
where the constant $L_0$ is the horizon value of the dilaton field determined from the full asymptotically flat string solution. $r_\pm$ are the outer and inner horizon radii and $\Omega_{{\rm AdS}_3}$ and $\Omega_{{\rm S}^3}$ are the volume forms of the unit-radius AdS$_3$ and S$^3$ respectively.
It has been checked \cite{BNOPT} that this form of the horizon geometry \eqref{NHS} is preserved under the inclusion of any curvature-squared super-invariant. The off-shell relation \eqref{sol1} also guarantees that the size of the AdS radius is set by the magnetic charge $P$. Thus, for the near-horizon geometry \eqref{NHS}, only $L_0$ is model-dependent. Below, we shall compute the BTZ black hole entropy encoding the central charges of the  asymptotic Virasoro symmetry algebra, assuming the Brown--Henneaux boundary condition, via
\be
S_{\rm BTZ}=\frac{\pi^2\ell}{3}\left(c_LT_L+c_RT_R\right)\,,\quad T_L=\frac{r_+-r_-}{2\pi\ell^2}\,,\quad T_R=\frac{r_++r_-}{2\pi\ell^2}\,.
\label{s2}
\ee
On the other hand, if the model enjoys a string theory origin, the central charges can also be computed microscopically using the $D=2$ CFT description of string theory in AdS$_3$. Comparison of the macroscopic and microscopic results then furnishes a precision test of the AdS$_3$/CFT$_2$ correspondence.

We begin with the most general $R+R^2$ action preserving ${\cal N}=(1,0)$ supersymmetry in $D=6$ \cite{BNOPT}
\be
S_{R+R^2}=\frac1{16\pi G_6}\int d^6x \left({\cal L}_{\rm EH}+\frac{1}{16}\l_{\rm GB}\,{\cal L}_{\rm GB}+\frac{1}{16}\l_{{\rm Riem}^2}\,{\cal L}_{{\rm Riem}^2}+\l_{R^2}\,{\cal L}_{R^2}\right)\,,
\label{IIAK3}
\ee
where the supersymmetric Riemann-squared action ${\cal L}_{{\rm Riem}^2}$ takes the form \cite{BSS86,BR87}
\be
{\cal L}_{{\rm Riem}^2}= \sr{-g} [ R_{\m\n\a\b}(\o_-) R^{\m\n\a\b}(\o_-)- \ft14 \e^{\m\n\r\s\l\t} B_{\m\n} R_{\r\s}{}^{\a}{}_{\b} (\o_-) R_{\l\t}{}^{\b}{}_{\a}(\o_-)+\ldots ] \,,
\ee
in which $R_{\m\n}{}^{\a}{}_{\b} (\o_-)$ is the curvature associated with the torsionful spin connection
\be
\o_{-\m}^{\a}{}_{\b}:=\o_{\m}^\a{}_\b-\ft12H_{\m}{}^\a{}_\b\,.
\ee
The last term in \eqref{IIAK3} denotes the supersymmetric Ricci scalar squared action whose explicit form can be found in \cite{BNOPT,O}. When auxiliary fields are switched off, which is the case considered here, ${\cal L}_{R^2}$ becomes the square of the dilaton field equation and thus does not contribute to on-shell quantities. We notice that the dilaton field only appears in ${\cal L}_{\rm EH}$, dropping out completely from the curvature-squared actions. This structure resembles the string theory low energy effective action with one-loop corrections. Indeed, if choosing
\be
\l_{\rm GB}=g_s^2\ell_s^2\,,\quad \l_{{\rm Riem}^2}=g_s^2\ell_s^2\,,
\label{coef}
\ee
the action \eqref{IIAK3} matches with the low energy effective action of type IIA string theory compactified on a K3 manifold \eqref{IIAK3} up to Ricci scalar squared terms that cannot be fixed by using string scattering amplitudes. K3 compactification of the type IIA string preserves sixteen supercharges, meaning that the particular combination with \eqref{coef} enhances the supersymmetry from ${\cal N}=(1,0)$ to ${\cal N}=(1,1)$. In our parametrization, the coefficients in front of the one-loop terms contain the string coupling $g_s$ because there is also a factor of $g_s^2$ hidden in the definition of $G_6$. The $D=6$ string coupling is determined by the asymptotic value of the dilaton field $L$.

BTZ black hole entropy in higher-derivative extended gravity models has been widely studied using the Iyer--Ward \cite{IW} formula and its generalizations concerning Chern--Simons type terms without manifest gauge invariance \cite{Tachikawa:2006sz,ALNR}.  In contrast to the non-supersymmetric models, supersymmetric completion of the Riemann tensor-squared necessary requires a $B\wedge R_\pm\wedge R_\pm$ term, which can also be recast as $H\wedge CS(\o_\pm)$. The former is manifestly invariant under coordinate transformations and local frame rotation, but transforms non-trivially under the one-form gauge symmetry of $B_{(2)}$.  The latter is exactly the opposite. These two different choices differing in a non-gauge invariant total derivative term yields different contributions to the black hole entropy. A na\"{\i}ve application of the Iyer--Ward \cite{IW} formula to the $B\wedge R_\pm\wedge R_\pm$ term produces a vanishing result as the resulting entropy formula is proportional to $R_{\m\n}{}^{\a}{}_{\b} (\o_\pm)$ vanishing on the solution \eqref{NHS}. However, using the Tachikawa formula \cite{Tachikawa:2006sz}, we do get a contribution proportional to $r_-$. The apparent ambiguity can be avoided by using an approach based on dimensional reduction. As the geometry is a product space of AdS$_3$ and S$^3$, it is more natural to use the three-dimensional effective action describing the dynamics in the non-compact AdS direction. Such an action should come from reducing the six-dimensional action on S$^3$. The validity of applying this approach is supported by computing the on-shell action of the AdS$_3\times$S$^3$ solution. One obtains the correct leading contributing from the $D=3$ Einstein--Hilbert action reduced from $D=6$, in contrast to the vanishing result obtained from the $D=6$ action. As for the contributions to black hole entropy from the parity-even part of the action \eqref{IIAK3}, the computation can still be done in $D=6$ or $D=3$ using the Iyer--Wald formula. The results are identical as the near-horizon geometry \eqref{NHS} solves the field equations derived from only the parity-even part of the action \eqref{IIAK3} and conserved charges are preserved by consistent dimensional reduction. Analogous to \cite{LPS}, in the reduction we retain the metric components in the three non-compact directions, the dilaton and the scalar field parametrizing the fluctuation of $H_{(3)}$ in the non-compact directions, which are sufficient to capture the BTZ$\times$S$^3$ solution \eqref{NHS}. The reduction of $B\wedge R\wedge R$ terms in \eqref{IIAK3} then give rise to
the Lorentz Chern--Simons action. Explicitly,
\be
-\frac{g_s^2\ell^2_s}{16\pi G_6}\int d^6x \sr{-g}\ft1{64}\left(\e^{\m\n\r\s\l\t} B_{\m\n} R_{\r\s}{}^{\a}{}_{\b} (\o_-) R_{\l\t}{}^{\b}{}_{\a}(\o_-)+\e^{\m\n\r\s\l\t} B_{\m\n} R_{\r\s}{}^{\a}{}_{\b} (\o_+) R_{\l\t}{}^{\b}{}_{\a}(\o_+)\right)
\ee
reduces to
\be
\frac{g_s^2\ell^2_s}{16\pi G_3\ell}I_{\rm CS}\,,\quad I_{\rm CS}=\frac{1}{2} \int d^3x \sr{-g} \epsilon^{\l\m\n}\Gamma^{\r}_{\l\s}( \partial_{\m}\Gamma^{\s}_{\r\n}
+\ft23\Gamma^{\s}_{\m\t}\Gamma^{\t}_{\n\r} ) \,,\quad \frac{1}{G_3}=\frac{2\pi^2\ell^3}{G_6}\,,
\label{podd}
\ee
to which the formula in \cite{Tachikawa:2006sz} can be applied straightforwardly.

To complete the stringy corrections to the BTZ black hole entropy, we also need the horizon value of the dilaton field in the dyonic string solution corrected by both the Gauss--Bonnet and the Riemann tensor squared term. The calculation resembles the one carried out in the previous section. We thus omit the details, but simply give the result
\be
L\vert_{r=0}=\frac{Q}P+\frac{\l_{\rm GB}+\l_{{\rm Riem}^2}}{2P}=\frac{Q}P+\frac{g_s^2\ell_s^2}{P}\,.
\ee
Summarizing contributions from various terms in the $R+R^2$ action \eqref{IIAK3}, we obtain the entropy of BTZ black boles embedded in the K3 compactification of the type IIA string,
\be
S_{\rm BTZ}=\frac{\pi^3\ell^3}{G_6}\left[ \bigg( \frac{Q}P+2\frac{g_s^2\ell_s^2}{\ell^2}\bigg) r_+ +\frac{g_s^2\ell_s^2}{\ell^2}r_-\right]\,.
\label{s1}
\ee
The Iyer--Wald formula derived from the 6D pure gravitational Gauss--Bonnet term vanishes when evaluated on the BTZ$\times$S$^3$ solution \eqref{NHS}. The additional contribution proportional to $r_+$ comes from the $RH^2$ terms that appear in the supersymmetric completion of the Gauss-Bonnet term. Comparing \eqref{s1} to \eqref{s2}, we read off the two central charges
\be
c_L=\frac{3\pi^2\ell^4}{G_6}\left(\frac{Q}{P}+\frac{g_s^2\ell_s^2}{\ell^2}\right)\,,\quad c_R=\frac{3\pi^2\ell^4}{G_6}\left(\frac{Q}{P}+3\frac{g_s^2\ell_s^2}{\ell^2}\right)\,.
\label{cc}
\ee

Upon lifting to type IIA string theory, the rotating string solution is composed of a bound state of $N_1$ fundamental strings and $N_5$ NS5-branes wrapped on K3. Therefore, the electric and magnetic energies carried by the asymptotically flat rotating string solution \eqref{genstr} should match with those of fundamental strings and NS5-branes. Using the fact that a NS5-brane wrapped on K3 becomes
a solitonic string in $D=6$ with effective tension
\be
T_{\rm NS5}{\rm Vol}(K3)\,,\quad T_{\rm NS5}=\frac{1}{(2\pi)^5g_s^2\ell_s^6}\,,\quad {\rm Vol}( K3)=(2\pi\ell_s)^4\,,
\ee
we obtain
\be
Q=\frac{2N_1G_6}{\pi^2\ell^2_s}\,,\qquad P=\frac{2N_5G_6}{\pi^2g_s^2\ell^2_s}\,.
\label{relation1}
\ee
On the other hand, the $D=6$ Newton's constant is related to the $D=10$ Newton constant $G_{10}$ via
\be
G_6{\rm Vol (K3)} =G_{10}\,,\qquad 16\pi G_{10}=(2\pi)^7g_s^2\ell_s^8\,.
\label{relation2}
\ee
Substituting \eqref{relation1}, \eqref{relation2} and $\ell^2=P$ into \eqref{cc}, we can express the central charges in terms of the quantized parameters
\be
c_L=6N_5(N_1+1)\,,\qquad c_R=6N_5(N_1+3)\,.
\label{cc2}
\ee
The geometric background is valid when the string is weakly coupled and the AdS radius is much larger than the string length, in other words, $g_s\ll1$ and $N_5\gg1$, under which the AdS radius is also much larger than the Planck length. The AdS$_3\times$S$^3\times$K3 vacuum in type IIA preserves eight chiral supercharges, which amounts into $(4,0)$ supersymmetry in $D=2$.  This means that the CFT$_2$ dual should be a $(4,0)$ SCFT with left and right $\widehat{SL(2)}\times\widehat{SU(2)}$ affine symmetry. The central charges of the dual CFT ought to match with those in the gravity side \eqref{cc2}, but we omit here an independent calculation of the central charges on the CFT side. Using the duality between the type IIA string on K3 and the heterotic string on $T^4$, we can map the solution obtained here to a solution in heterotic string theory with leading order stringy corrections. Denoting the fields in the heterotic string by $\{\widetilde{g}_{\m\n}\,,\widetilde{B}_{\m\n}\,,\widetilde{L} \}$, the $D=6$ type IIA/heterotic string duality at ${\cal O}(\ell_s^2)$ is given by \cite{LPS, LM}
\be
\widetilde{H}_{(3)}=L*H_{(3)} \,,\qquad \widetilde{g}_{\m\n}=L g_{\m\n}\,,\qquad \widetilde{L}=L^{-1}\,,
\ee
under which the low energy effective action of the type IIA string with leading one-loop corrections \eqref{IIAK3} is mapped to the effective action of the heterotic string with leading tree-level stringy corrections
\be
\widetilde{S}=\frac1{16\pi G_6}\int d^6x\sr{-\widetilde{g}}\widetilde{L}[\widetilde{R}+\widetilde{L}^{-1}\nabla^\mu \widetilde{L} \, \nabla_\mu \widetilde{L} - \ft{1}{12}\widetilde{H}^{\m \n \r} \widetilde{H}_{\m\n\r}+\ft18\ell_s^2\widetilde{R}_{\m\n\a\b}(\o_+) \widetilde{R}^{\m\n\a\b}(\o_+)+\ldots]
\ee
in which $\widetilde{H}_{(3)}$ obeys the non-trivial Bianchi identity
\be
d\widetilde{H}=\ft14\ell_s^2{\rm tr}(\widetilde{R}_+\wedge \widetilde{R}_+)+\ldots\,.
\ee
In the formulae above, we have omitted the spin-1 gauge fields, which are not relevant here. Holographic central charges for the dualized AdS$_3\times$S$^3$ solution can be readily computed by noticing a few shortcuts. Firstly, the contribution from the
Einstein--Hilbert action is unchanged because the Einstein-frame metric remains the same under the duality transformation. Secondly, the parity-odd contribution is opposite to that of \eqref{IIAK3}, since the on-shell relation $\widetilde{L}*\widetilde{H}=H$ means that there is effectively an $H\wedge {\rm CS}(\o_+)$ term with coefficient $-\ft14\ell_s^2$. Finally,
the parity-even $\widetilde{R}_+^2$ term does not contribute, since $\widetilde{R}_+$ vanishes for the solution. We thus obtain the entropy of BTZ black boles embedded in the $T^4$ compactification of the heterotic string
\be
\widetilde{S}_{\rm BTZ}=\frac{\pi^3\ell^3}{G_6}\left[ \bigg( \frac{Q}P+\frac{g_s^2\ell_s^2}{\ell^2} \bigg) r_+-\frac{g_s^2\ell_s^2}{\ell^2}r_-\right]\,.
\ee
It is evident that the duality transformation interchanges the role of the fundamental and solitonic strings. Thus in the dualized solution, $Q$ is related to $\widetilde{N}_5$, the number of heterotic NS5-branes, while $P$ is related to $\widetilde{N}_1$, the number of heterotic fundamental strings. In terms of the quantized parameters, the central charges for the
AdS$_3\times$S$^3\times T^4$ solution in the heterotic string are expressed as
\be
\widetilde{c}_L=6\widetilde{N}_1(\widetilde{N}_5+2)\,,\qquad \widetilde{c}_R=6\widetilde{N}_1\widetilde{N}_5\,.
\label{cc2b}
\ee
The CFT dual to the heterotic string in the AdS$_3\times$S$^3\times T^4$ background was studied in \cite{KLL}. It is a $(4,0)$ SCFT with left and right $\widehat{SL(2)}\times\widehat{SU(2)}\times\widehat{U(1)}^4$ affine symmetry.
The central charges in the left and right movers are
\be
c_L=6p(k+2)\,,\qquad c_R=6pk\,,
\label{cc3}
\ee
where $p$ is the number of fundamental strings and $k$ is the level of the $\widehat{SL(2)}$ algebra in the supersymmetric right mover. The matching between \eqref{cc2b} and \eqref{cc3} leads to the identification
\be
k=\widetilde{N}_5\,.
\label{id}
\ee

\section{Conclusion and discussion}

In this paper, we proposed a method for constructing BPS
solutions in higher-derivative extended supergravity models from known solutions in two-derivative theories. Application of our method requires an off-shell formulation underlying the supergravity theory. For simplicity, we focused on 6D ungauged supergravity models involving only the dilaton--Weyl multiplet. Adding off-shell vector multiplets to the model is straightforward.  A concrete example was given by obtaining the first rotating dyonic string solution in 6D Einstein-Gauss-Bonnet supergravity, exhibiting the advantage of our method when applied to non-static solutions. An obvious generalization of our solution
is to turn on monopole charge $M$ on $S^3$. In this case, the solution respects the same U(2)$\times\mathbb{R}^2$ symmetry and remains cohomogeneity-1. The entropy of the BTZ black hole arising from the near-horizon limit of the dyonic string in monopole background will be labeled by three integers \cite{KLL}. Upon dualizing the solution from type IIA string theory to the heterotic string, we will be able to compare the gravity and dual CFT results \cite{KLL}. We would also like to apply our construction to the more intricate cohomogeneity-2 case, in particular to the smooth horizonless microstate geometries \cite{Bena:2016ypk, Bena:2017geu} that have the same asymptotic structure at infinity as a given supersymmetric black hole of the same mass, charges and angular momenta. The microstate geometries also admit an AdS$_3\times$S$^3$ throat. It would be interesting to investigate if the throat region of the microstate geometry receives the same stringy corrections as the near-horizon limit of the dyonic strings. Finally, the off-shell supersymmetric curvature-squared terms are also known in $D=4$ and $D=5$. Their component forms are explicitly given in \cite{Cecotti:1986jy}-\cite{OP2}. Some of them do not come from dimensional reduction of the 6D theories and require separate consideration. Although many stationary BPS solutions have already been discovered in the corresponding two-derivative theories, the known supersymmetric black holes in $D=4$ and $D=5$ higher-derivative extended supergravities are restricted to the static case (see e.g.\ \cite{Behrndt:1998eq,Castro:2008ne}), so there remains much to be done.

\section*{Acknowledgements}
	We are grateful to T.\ Azeyanagi, G.\ Comp\`{e}re, M.\ Duff, H.\ L\"{u}, S.\ Schafer-Nameki and J.\ Sparks for useful discussions. We also thank D.\ Kutasov and F.\ Larsen for useful correspondence. It is also a pleasure to thank D.\ Butter, J.\ Novak, M.\ Ozkan and G.\ Tartaglino-Mazzucchelli for collaborations on related projects. The work of Y.P.\ is supported by the Newton International Fellowship NF170385 of the Royal Society.
	

\end{document}